# Improving the Secrecy of Distributed Storage Systems using Interference Alignment

Natasa Paunkoska, The University of Information Science and Technolog "St. Paul the Apostle", Ohrid, Macedonia
Ninoslav Marina, The University of Information Science and Technolog "St. Paul the Apostle", Ohrid, Macedonia
Venceslav Kafedziski, Faculty of Electrical Engineering and Information Technologies, Skopje, Macedonia

**ABSTRACT**
Regenerating codes based on the approach of interference alignment for wireless interference channel achieve the cut-set bound for distributed storage systems. These codes provide data reliability, and perform efficient exact node repair when some node fails. Interference alignment as a concept is especially important to improve the repair efficiency of a failed node in a minimum storage regenerating (MSR) code. In addition it can improve the stored data security in presence of passive intruders. In this paper we construct a new code resilient against a threat model where a passive eavesdropper can access the data stored on a subset of nodes and the downloaded data during the repair process of a subset of failed nodes. We achieve an optimal secrecy capacity for the new explicit construction of MSR interference alignment code. Hence, we show that the eavesdropper obtains zero information from the original message stored across the distributed storage, and that we achieve a perfect secrecy.

## 1  INTRODUCTION

Distributed storage systems (DSS) allow distributed storage of data file (message) in all new generation applications and provide more possibilities to improve its reliability and security. The data is stored in a decentralized manner across several unreliable distributed servers (nodes) in the system. The data distribution principle of the current storage systems is a simple scheme of triple replication of each file piece stored on some node. The replication of each piece must be placed in a node different from the node where the original file is stored. This practice is necessary, since one of the most common challenges in DSS, is the so called *repair process*, which happens when some nodes in the system fail and lose data. In that case, some of the replication data files of the failed original piece are used to recover the lost data. Note that this method is suboptimal in terms of bandwidth. In (Dimakis, 2010), Dimakis et al. (2010) found a new way of introducing error correcting codes to achieve better efficiency in the storage networks. They called these codes *regenerating codes*. They are efficient with respect to the storage utilization and the amount downloaded for repair. Different coding techniques could be combined to achieve various goals, such as minimizing the storage in each node, or minimizing the amount of downloaded data needed to repair the failed data. These two goals cannot be achieved at the same time, so one has to make a compromise between these two points called the minimum storage regeneration (MSR) point and the minimum bandwidth regeneration (MBR) point, respectively. The description of the region of achievable rates bounded by MSR and MBR was studied extensively in the previous years (Rashmi, 2012; Wu, 2007; Dimakis, 2006; Shah, 2010; Cadambe, 2011).

Let $B$ be the size of the message that needs to be stored in the system. This message is divided into pieces and stored in all $n$ nodes in the network. The size of each piece of the message stored in a single node is $\alpha = \frac{B}{k}$. Here, $k$ $(k<n)$ is the number of nodes that are going to be contacted by the data collector (user) for reconstructing the original message. Another parameter is $\beta$ $(\beta < \alpha)$, the amount of downloaded data from a single node during the repair process. The number of contacted nodes for accomplishing the repair process for recovering the lost data is $d$ $(k \leq d < n)$.

The parameters in such a regenerating code that aims to reliably store the file of a maximum size $B$ and achieve the cut-set bound in DSS, examined in (Dimakis, 2010), must satisfy the following condition

$$B \leq \sum_{i=0}^{k-1} \min\{\alpha, (d-i)\beta\}. \tag{1}$$

Based on the tradeoff between $\alpha$ and $\gamma = d\beta$ the two extreme points can be obtained as presented in (Wu, 2007). The paper describes minimization of both $\alpha$ and $\beta$ parameters. Minimizing $\alpha$ results in a minimum storage solution, while minimizing $\beta$ (for fixed $d$) results in a storage solution that minimizes repair bandwidth. From (1) can be concluded that is not possible to minimize both $\alpha$ and $\beta$ simultaneously and that there is tradeoff between the parameters $\alpha$ and $\beta$. Thus, in the reconstruction case when the storage per node $\alpha$ achieves its minimum $\frac{B}{k}$, the coding scheme achieves the extreme point MSR, and it is given by

$$(\alpha_S, \gamma_S) = \left(\frac{B}{k}, \frac{B}{k} \frac{d}{d-k+1}\right). \tag{2}$$

Otherwise, in the case when the repair bandwidth $d\beta$ is equal to $\alpha$, is achieved the extreme point MBR, given by the pair

$$(\alpha_B, \gamma_B) = \left(\frac{B}{k} \frac{2d}{2d-k+1}, \frac{B}{k} \frac{2d}{2d-k+1}\right). \tag{3}$$

In this paper we focus on the MSR point. The used concept includes the Interference Alignment method that is widely used for improving the capacity in the wireless communication [and JafarCadambe and Jafar2008], and here is adapted to adjust for the DSS and to perform very efficient repair process when some node fail (Shah, 2012).

Besides the reliability and the availability in storage networks, the security appears to be an additional challenge. A distributed data storage system is formed by many nodes widely spread across the Internet. So, each node in such a peer-to-peer network is vulnerable and a potential point of attack. The attackers can eavesdrop the nodes and possibly modify their data. The storage systems distinguish two types of passive attacks or observations. In the first type of attack, the eavesdropper observes the data stored on a subset of nodes in the system, and in the second type of attack the eavesdropper observes all the downloaded data during the repair process of a subset of the new nodes. This work aims at providing an explicit construction of a regenerating code that uses the interference alignment method, which achieves perfect secrecy. The paper is organized as follows. In Section 2, we describe the use of the interference alignment method in the distributed storage context. In Section 3, we give the general approach for providing security of the code construction. Section 4 proves the perfect secrecy of the new secure code construction together with some performance analysis and the paper concludes in Section 5.

## 2  INTERFFERENCE ALIGNMENT METHOD IN DSS

Code construction for data storage in distributed systems is based on the principle of interference alignment (IA) method. The code construction is for a regenerating code that achieves the cut-set bound. The idea of IA comes from wireless communications, aimed to design the signals of multiple users in such a way that at every receiver, signals from all the unintended users occupy a subspace of the given space, leaving the remainder of the space free for the signal of the intended user, as explained in (Cadambe & Jafar, 2008). This improves the degree of freedom that represent the rate of growth of network capacity with the log of the signal to noise ratio (SNR).

In the distributed storage context, the IA method is applied during the exact repair process of a failed node in a minimum storage regenerating (MSR) code. The construction by using this method is done by considering a $(n,k)$ systematic code, where the first $k$ nodes are an systematic and, thus, they store $k$ (uncoded) independent symbols. The remaining $(n-k)$ nodes are parity nodes. Linear combinations of the $k$ symbols, where the combinations are defined by the code generation matrix, are stored in these nodes. When there is a failed node, the newcomer (new node) downloads a certain linear combination of the information stored at each of the $(n-1)$ surviving nodes. The goal is to recover the lost data from the failed node using that set of linear combinations. Assuming that a systematic node has failed, from the $(k-1)$ surviving nodes the newcomer receives the uncoded independent symbols. The information from the failed node is stored in the $(n-k)$ parity nodes, noting that the information is mixed with the remaining $(k-1)$ symbols from the $(k-1)$ systematic nodes. These $(k-1)$ symbols which are not required by the new node, but arrive in the linear combinations downloaded from the parity nodes, due to the fact they are mixed with the failed symbols, are analogous to the interference in the wireless communication systems. The coding matrices used for making the combinations in the parity nodes, are analogous to the channel matrices in the wireless communications that perform the same function. In the repair process, the downloaded combinations by the newcomer are analogous to the beamforming vectors in wireless communications, elaborated in (Suh & Ramchandran, 2010). In the context of wireless communications, interference alignment reduces the footprint of the interference at the receiver and enables a greater number of dimensions for the desired signal. In the context of repair, interference alignment reduces the footprint of the interfering symbols at the newcomer, which means a smaller number of units to be downloaded to cancel the interference. However, one important note is that the channel matrices in wireless communications are given by nature and cannot be controlled, while in the storage systems, the coding matrices are a design choice.
In (Shah, 2012), the construction of an MSR code by using the interference alignment method is explained, where the Cauchy matrix is used as a coding matrix. A detailed explanation is given regarding the repair process, when the failed nodes are only the systematic nodes, only the parity nodes, and a combination of the two, and of the data reconstruction process. By our point of view we are investigating the level of security that can be achieved in the distributed storage systems using this way of data distribution.

## 3   APPROACH FOR PROVING SECRECY

Security is quite important aspect in a distributed storage network. The threat model is such that the eavesdropper may gain access to the data stored in a subset of the storage nodes, and also, to the data downloaded during the repair process of some other subset of nodes. Explicit construction of regenerating codes that achieve information-theoretic secrecy is provided in (Shah, 2011; Rawat, 2014). The principle of enabling security is based on the Wiretap channel II described in (Ozarow & Wyner, 1985). The main goal is to construct secure $[n,k,d]$ code to achieve MSR using interference alignment code given in (Shah, 2012). We denote the number of message symbols that can be securely stored in a distributed system as $B^{(s)}$. As an input to the MSR interference alignment code, when there is no secrecy, we need to choose a set of message symbols that will be replaced with random symbols $R$ chosen uniformly and independently from the finite field $F_q$ over which the code is defined, where

$$R = B - B^{(s)}.$$

The secure code will be identical with the original code, if we treat the random symbols as message symbols. To prove secrecy in this code construction we first need to consider the worst case scenario, the threat model, where the eavesdropper has access to $l_1$ (data stored on a subset of nodes), where the set of eavesdropped indices is denoted by $E_1$ and $l_2$ (data downloaded during repair of $l_2$ nodes) nodes, where the set of eavesdropped indices is denoted by $E_2$. The total number of compromised nodes in the distributed system can not be greater than $k$, or $(l = l_1 + l_2) < k$. Moreover, $l_1$ and $l_2$ are disjunctive nodes. The proof of the information-theoretic secrecy of this code is established as follows:

- Step 1: Show that given the collection of the $B^{(s)}$ secure message symbols $U$ as side information, the eavesdropper can recover all $R$ random symbols i.e., $H(R|\varepsilon,U) = 0$. ($R$ denotes a collection of $R = B - B^{(s)}$ random symbols, and $\varepsilon$ is a collection of symbols that the eavesdropper gains access to).
- Step 2: Show that all but $R$ of the symbols obtained by the eavesdropper are functions of these $R$ symbols, i.e., $H(\varepsilon) \leq H(R)$.
- Step 3: Show that the two conditions in steps 1 and 2 necessarily imply that the mutual information between the message symbols $U$ and the symbols obtained by the eavesdropper $\varepsilon$, is zero, i.e., $I(U;\varepsilon) = 0$.

## 4   INFORMATION-THEORETIC SECRECY IN INTERFERENCE ALIGENMENT METHOD IN DSS

Pawar et al. in (Pawar, 2010) provide an upper bound on $B^{(s)}$, given by

$$B^{(s)} \leq \sum_{i=l}^{k-1} \min(\alpha, (d-i)\beta). \tag{4}$$

The interpretation of the bound in (4) is that out of $k$ nodes to which a data collector connects, the first $l$ of these nodes are compromised. Thus, by the assumption that the secrecy goals have been met, these $l$ nodes will not provide any information about the message symbols. Only the remaining $(k-l)$ nodes may provide useful information.

From the MSR point of view the repair bandwidth is strictly greater than the per node storage and an eavesdropper potentially obtains more information when she has an access to the data downloaded during the node repair process. Therefore, for the MSR point of view from (4), the security upper bound becomes,

$$B^{(s)} \leq (k-l)\alpha. \tag{5}$$

For the constructed code in this paper, the following result is stated: Assume code construction for $n=2k$ and $d=n-1$, in the case of $(l_1, l_2)$ eavesdropper model, where $(l = l_1 + l_2) < k$, the MSR secure bound in (5), when interference alignment is used becomes

$$B^{(s)} = (k - l_1 - l_2)(\alpha - l_2). \tag{6}$$

*Proof.* This equality holds for all $[n=2k, k, d=n-1]$ MSR codes, where $B = k\alpha$, $d\beta = \alpha + (k-1)\beta$, $\alpha = k$ and $\beta = 1$ given in (Shah, 2011). The total number of symbols that can be stored in MSR distributed system is $B = k\alpha$. From these symbols we need to subtract all the compromised symbols. From the first type of observation total number of eavesdropped symbols is $l_1 \alpha$, the observed $l_1$ nodes will not provide any useful information. Plus from the second type of observation there are $(n-l_2)\beta l_2$ downloaded symbols during the repair process minus these already known from the first type of observation $\beta l_1 l_2$. Thus, the formulation of (6) is following,

$$\alpha k - (l_1 \alpha + (n - l_2)\beta l_2 - \beta l_1 l_2) =$$
$$\alpha k - (l_1 \alpha + 2k\beta l_2 - l_2^2 \beta - \beta l_1 l_2) =$$
$$\alpha k - l_1 \alpha - k\beta l_2 - k\beta l_2 + l_2^2 \beta + \beta l_1 l_2 =$$
$$k^2 - kl_1 - kl_2 - kl_2 + l_2^2 + l_1 l_2 =$$
$$k(k - l_1 - l_2) - l_2(k - l_1 - l_2) =$$
$$(k - l_2)(k - l_1 - l_2) =$$
$$(k - l_1 - l_2)(\alpha - l_2)$$

Now we need to construct a secure MSR code based on the Interference Alignment concept, using the eavesdropper model $(l_1, l_2)$ that will satisfy the equality (6). As described in (Shah, 2012), the output codewords will be denoted by $C = uG^{(m)}$, where $u$ is the message of size $B$ and the secure MSR interference alignment code will be $C^{(s)} = u_n^s G^{(m)}$, where $u_n^s$ is new message consist of $u^s$ secure message of size $B^{(s)}$ plus $R$ random symbols. The construction will be made for the case when $\beta = 1$, $\alpha = d - k + 1 = k$. The secure MSR interference alignment code need to be constructed from the modified original message $u_n^s$, consisting of $k\alpha$ message symbols from which $R = B - B^{(s)}$ symbols will be replaced with random variables, and a generator matrix $G^{(m)}$, for $m = 1,...,n$. $G^{(m)}$ in (Shah, 2012) is defined as,

$$G^{(m)} = \begin{bmatrix} G_1^{(m)} \\ \vdots \\ G_k^{(m)} \end{bmatrix},$$

where $m = 1,...,k$ are the systematic nodes, and $m = k+1,...,n$ are the parity nodes. The parity generator matrices include a Cauchy matrix (Bernstein, 2005), a matrix with special construction that efficiently performs the repair and reconstruction process.

### 4.1 GENERAL SECURE CODE CONSTRUCTION

In this subsection, we present the general construction of a coding scheme that is secure against an $(l_1, l_2)$ eavesdropper when $E_1 \cup E_2 \subseteq E$, for a given set $E$ with cardinality $|E| < k$, for all parameter values $[n = 2k, k, d = n-1]$ and $d = \alpha + k - 1$, $\beta = 1$. The code properties indicate that $k = \alpha$, which has the main role during the process of designing generator matrices for the parity nodes. This relation ensures that each node reserve $\alpha = k$ symbols with linearly independent global kernels used for repair of $k$ systematic nodes. The construction is based on MSR interference alignment code where the message content is modified by $B - B^{(s)}$ random symbols.

The construction will depend based on the following properties associated with the repair process in a MSR interference alignment code:

**Lemma 1.** *Assume that an eavesdropper gains access to the data stored on $l = l_1$ nodes in an MSR interference alignment code. Then, the eavesdropper can only observe $l\alpha$ independent symbols.*

*Proof.* Since the size of the stored data on each node is $\alpha = k$ symbols by the construction of MSR interference alignment code the maximum number of independent symbols that the intruder can reveal is $lk$ if $E \subset [k]$.

**Lemma 2.** *Assume that an eavesdropper has an access to the data stored on any $l_1$ nodes and observes the downloaded data from $l_2$ nodes that are in reparation process in an MSR interference alignment code. Then the $(l_1, l_2)$ eavesdrooper can observe at most $l_1\alpha + (n - l_2)\beta l_2 - \beta l_1 l_2$ independent symbols.*

*Proof.* From Lemma 2 in an $(l_1, l_2)$ eavesdrooper model, $(l_1 + l_2) < k$ the intruder can only observe $kl_1$ independent symbols when it gains access in the data stored on $l_1$ nodes. Since in MSR interference alignment code the repair bandwidth is $\beta d$, where $d = n-1$, $\beta = 1$, i.e., a newcomer node can recover the lost symbols stored in the failed node by downloading a single symbol from any $d$ nodes, the maximum number of independent symbols that the intruder can reveal is $(n - l_2)\beta l_2$ if $E_2 \subset N$, $N$ ($|N| \leq k$) set of indices of systematic nodes, when it gains access the downloaded data of $l_2$ failed systematic nodes. Therefore, the maximum number of message symbols that the intruder can reveal if it can read-access the data stored in $l_1$ nodes and read-access the downloaded data during the repair process of $l_2$ failed systematic nodes is $l_1\alpha + (n-l_2)\beta - \beta l_1 l_2$, where $\beta l_1 l_2$ are already observed symbols from the first observation.

**Definition 1** *(Cauchy Matrix (Bernstein, 2005)):* An $(s \times t)$ Cauchy matrix $\Psi$ over a finite field $\mathsf{F}_q$ is matrix whose $(i, j)$-th element $(a \leq i \leq s, 1 \leq j \leq t)$ equals $\dfrac{1}{(x_i - y_i)}$ where $\{x_i\} \cup \{y_i\}$ is an injective sequence, i.e., a sequence with no repeated elements.

For the construction of a $(s \times t)$ Cauchy matrix the minimum field size will be $s + t$. Thus by choosing $\Psi$ to be a Cauchy matrix, leads to

$$q \geq \alpha + n - k. \tag{7}$$

For the concrete construction any finite field that satisfies this condition will be sufficient, since $n - k \geq \alpha \geq 2$, and $q \geq 4$.

For constructing a new secure code based on an MSR interference alignment method we consider the secure message $u^s$ of size $B^{(s)}$ over $\mathsf{F}_q$, i.e., $u^s = (a_1, a_2, ..., a_{(k-l_1-l_2)(\alpha-l_2)})$. From there we take $(l_1 + l_2)\alpha + (k - l_1 - l_2)l_2$ i.i.d. random symbols $r = (r_1, ..., r_{(l_1+l_2)\alpha + (k-l_1-l_2)l_2})$ distributed uniformly at random over $\mathsf{F}_q$. The set of random symbols $r$ is appended to the secure message to obtain the general message $u_n^s = (r, u^s) \in \mathsf{F}_q$, that will be encoded in the following manner:

The design of the achivability scheme for MSR interference alignment code based on (Shah, 2012) is folowing:

- Design of Systematic Generator Matrices:

In an MSR interference alignment code the first $k$ nodes are systematic and the message symbols that are stored there are in uncoded form. Therefore, the generator matrices for those nodes $G_i^{(\ell)}$, $1 \leq i \leq k$, of the $\ell$-th systematic node, $1 \leq \ell \leq k$, will consist of

$$G_i^{(\ell)} = \begin{cases} I_\alpha & \text{if } i = \ell \\ 0_\alpha & \text{if } i \neq \ell. \end{cases} \tag{8}$$

- Design of Parity Generatior Matrices:

For designing of the parity matrices it must be chosen matrix $\Psi$ with dimension $\alpha \times (n-k)$ with entries drawn from $\mathsf{F}_q$ such that every submatrix of $\Psi$ has full rank. In this construction, because $n - k = \alpha = k$, $\Psi$ is a square matrix. The columns of $\Psi$ are defined by

$$\Psi = [\underline{\psi}^{(k+1)} \quad \underline{\psi}^{(k+2)} \quad \cdots \quad \underline{\psi}^{(n)}] \tag{9}$$

where the $m$-th column is given by

$$\underline{\psi}^{(m)} = \begin{bmatrix} \psi_1^{(m)} \\ \vdots \\ \psi_\alpha^{(m)} \end{bmatrix}. \tag{10}$$

The matrix $\Psi$ that is used for constructing parity matrices and meets the criteria for the repair process is the so-called Cauchy matrix (Shah, 2012).

For better clarification a new notation is introduced, i.e., the $j$-th column of the $(\alpha \times \alpha)$ matrix $G_i^{(m)}$ is denoted as $\underline{q}_{i,j}^{(m)}$, or

$$G_i^{(m)} = [\underline{q}_{i,1}^{(m)} \ldots \underline{q}_{i,\alpha}^{(m)}]. \tag{11}$$

The Interference Alignment algorithm is designed such that each of the $\alpha$ parity nodes in a case of the repair process of the $\ell$-th systematic node will pass its $\ell$-th column. Regarding the above statement, for the parity nodes $k+1 \leq m \leq n$, $1 \leq i, j \leq \alpha$, we select

$$\underline{q}_{i,j}^{(m)} = \begin{cases} \varepsilon \underline{\psi}^{(m)} & \text{if } i = j \\ \psi_i^{(m)} \underline{e}_j & \text{if } i \neq j, \end{cases} \tag{12}$$

where $\varepsilon$ is an element from $\mathbb{F}_q$ so that $\varepsilon \neq 0$ and $\varepsilon^2 \neq 1$ (needed during the reconstruction process). Note that $\varepsilon$ exists as long as $q \geq 4$. This property is required to have a successful repair process in the DSS code construction.

The design of this code is made to be in line with the properties and conditions of the wireless interference concept, which is particularly significant in the exact repair of the systematic nodes. Therefore, after the design of the generator matrices the new general secure Interference Alignment code construction is made by multiplication of the secure message $u^s$ and the generator matrix $G^{(m)}$,

$$C^{(s)} = u_n^s G^{(m)}. \tag{13}$$

Using the newly constructed secure MSR interference alignment code, we are able to prove the properties that satisfy the reconstruction, repair and security processes.

**Theorem 1.** *(Reconstruction process): The data collector can reconstruct all $B^{(s)}$ message symbols of the newly constructed $C^{(s)}$ code by contacting any $k$ nodes in the system.*

*Proof.* The sketch of the proof is following, treating the random symbols as message symbols the secure MSR interference alignment code $C^{(s)}$ becomes identical to the MSR interference alignment code $C$ given in (Shah, 2012). During the reconstruction process, the data collector contacts any $k$ nodes in the system by downloading $\alpha$ symbols from each of them. In total, the amount of information that it gets is $k\alpha$. In case of all connected nodes to be systematic, which are pure symbols, the obtaining of the original message is straight and easy. In a case of combination of them or all $k$ nodes to be parity, first it need to be cancelled all interfered symbols and after that is possible to be extracted the wanted message. The readers can check this part using (Shah, 2012).

**Theorem 2.** *(Repair process): During the repair process, the newcomer can recover the failed data in the newly constructed $C^{(s)}$ code by contacting $d$, number of contacted nodes for performing the repair process for recovering the lost data, nodes in the system.*

*Proof.* Same as for Theorem 1, for Theorem 2 we consider that the random variables can be seen as message symbols, and then the secure MSR interference alignment code $C^{(s)}$ becomes identical to the MSR interference alignment code $C$ elaborated in (Shah, 2012), where repair processes is possible. The sketch of the proof is that when some node fail, a newcomer comes in the system and contacts any $d$ live nodes by downloading from them only a single symbol (component) of the stored $\alpha$ information. The total downloaded information is $d = n-1 = 2k-1$, where $k = \alpha$, and we only need to repair the $\alpha$ lost data. For this reason, interference cancellation will be performed leaving only part of the downloaded data. After that the desired component is obtained by multiplying the rest of data with the appropriate inverse Cauchy submatrix. The proof given in (Shah, 2012) is adequate to be applied in the same manner by the new created secure MSR interference alignment code $C^{(s)}$ in this paper. Because of preserving space and similarity of the proof, the readers can check this part using (Shah, 2012).

**Theorem 3.** *(Information-theoretic secrecy): In the code $C^{(s)}$ designed to be secure against a threat model $(l_1, l_2)$, where the eavesdropper has an access to the data stored in $l_1$ nodes and downloaded data from $l_2$ nodes, the eavesdropper is not able to obtain any information for the original message, i.e., $I(U; \varepsilon) = 0$.*

*Proof.* Let the eavesdropper in the first observation have access to $l_1 k$ stored data in the system, in our case $k = \alpha$. This means she has an access to $B_{s(1)}^{eve} = l_1 k$ symbols from total $B = k\alpha$, forming the message $u_s^{eve} = [r_1, \ldots r_k, 0_{k+1}, \ldots, 0_{k\alpha}]$, in the $C^{(s)}$ code. The eavesdropper, knowing the coding scheme, from the first observation will get the following construction, $E_1^{(s)} = u_s^{eve} G^{(m)}$. Additionally, another type of observation is accessing the downloaded data equal to $B_{s(2)}^{eve} = d\beta l_2 = d l_2$ symbols during the repair process, where $\beta = 1$ in our construction. Thus by Theorem 2 in (Shah, 2012), in the worst case scenario when a failed node is a systematic node, it can be exactly repaired. The same node can be accessed by the eavesdropper and she can download one symbol from each of the remaining nodes. If we consider that the systematic node $\ell$ fails, each of the remaining $d = n-1$ nodes will pass its $\ell$-th column, so that the eavesdropper in the second observation obtains the following columns during the repair process,

$$E_2^{(s)} = \begin{bmatrix} \underline{e}_\ell & \cdots & \underline{0} & \underline{0} & \cdots & \underline{0} & \psi_1^{(k+1)} \underline{e}_\ell & \cdots & \psi_1^{(n)} \underline{e}_\ell \\ \vdots & \ddots & \vdots & \vdots & \ddots & \vdots & \vdots & \ddots & \vdots \\ \underline{0} & \cdots & \underline{e}_\ell & \underline{0} & \cdots & \underline{0} & \psi_{\ell-1}^{(k+1)} \underline{e}_\ell & \cdots & \psi_{\ell-1}^{(n)} \underline{e}_\ell \\ \underline{0} & \cdots & \underline{0} & \underline{0} & \cdots & \underline{0} & \varepsilon \psi^{(k+1)} & \cdots & \varepsilon \psi^{(n)} \\ \underline{0} & \cdots & \underline{0} & \underline{e}_\ell & \cdots & \underline{0} & \psi_{\ell+1}^{(k+1)} \underline{e}_\ell & \cdots & \psi_{\ell+1}^{(n)} \underline{e}_\ell \\ \vdots & \ddots & \vdots & \vdots & \ddots & \vdots & \vdots & \ddots & \vdots \\ \underline{0} & \cdots & \underline{0} & \underline{0} & \cdots & \underline{e}_\ell & \psi_k^{(k+1)} \underline{e}_\ell & \cdots & \psi_k^{(n)} \underline{e}_\ell \end{bmatrix},$$

where $\underline{e}_\ell$ denotes the $\ell$-th unit vector of length $\alpha$ and $\underline{0}$ denotes a zero vector of length $\alpha$. The element $\varepsilon$ from $\mathsf{F}_q$ is $\varepsilon \neq 0$ and $\varepsilon^2 \neq 1$, by Definition 1, which is necessary condition for performing the reconstruction process.

Except for the $\ell$-th component in the matrix, every other component is aligned with the vector $\underline{e}_\ell$, which will show that some $\alpha$ linear combinations of the columns above will give us a matrix whose $\ell$-th component equals the $(\alpha \times \alpha)$ identity matrix, and has zeros everywhere else.

This is a consequence of the interference alignment structure in combination with the linear independence of the $\alpha$ vectors in the desired component,

$$\{\underline{\psi}^{(k+1)},...,\underline{\psi}^{(n)}\},$$

where the $m$-th column is given by

$$\underline{\psi}^{(m)} = \begin{bmatrix} \psi_1^{(m)} \\ \vdots \\ \psi_\alpha^{(m)} \end{bmatrix}.$$

The eavesdropper after both constructions $E_1^{(s)}$ and $E_2^{(s)}$ can examine which symbols have been revealed. The symbols which are obtained by eavesdropping are gain from two different approaches, meaning that we can conclude that some of the symbols are repeated. The number of repeated symbols $h_r$ should be subtracted from the total observed symbols, i.e., these symbols should not be taken into account. Regarding this, the total number of message symbols observed by the eavesdropper is obtained as the sum of the two types of observations explained above, and are given by

$$B_s^{eve} = B_{s(1)}^{eve} + B_{s(2)}^{eve} - h_r.$$

According to the Step 1 for proving information-theoretic secrecy, first we need to show that for given message symbols as a side information, the eavesdropper can decode all the random symbols. Therefore, for the first type of observation we generate a new message vector $u_s^{eve}$ of length $k\alpha$ consisting of only the symbols seen by the eavesdropper, and all other symbols zeros. This part of the side information is defined as,

$$\tilde{E}_1^{(s)} = u_s^{eve} G^{(m)},$$

where $m = 1,2,...,k$ are the systematic nodes and $m = k+1,...,n$ are the parity nodes.

Our construction is performed for the worst case scenario, or when $l = l_1 + l_2 = k - 1$, by the assumption that the number of affected nodes is always smaller then the number of systematic nodes $l < k$. Therefore, for $l_1$ observed nodes the new message will be $u_s^{eve} = [r_1,...r_k,0_{k+1},...,0_{k\alpha}]$ and $\tilde{E}_1^{(s)}$ in the construction will have only information for these symbols. We are assuming that the eavesdropper knows the encoding scheme.

For the second type of observation, we take the message $u_n^s$ and define a new generator matrix $G_{eve}^{(m)}$. The matrices $G_{eve}^{(m)}$ for the $m = 1,...,k$, systematic nodes, accordingly will have non zero elements on the places where the eavesdropper has access during the repair and zeros elsewhere. Similarly is done also the design of new generator matrix $G_{eve}^{(m)}$ for the parity nodes $m = k+1,...,n$, since we assume that the failed node is a systematic node that corresponds to the $l_2$ nodes in which the eavesdropper can access during the repair downloads. For this case, when node $\ell$ fails the $\ell$-th column in every matrix except the failed one will have non zero element, and in the other columns zeros. This means that the Cauchy matrices in the generator matrix have nonzero columns only at the places where the eavesdropper has access (parts that will be downloaded during the repair according to which nodes have failed), so that the second part of the side information in our particular construction will be

$$\tilde{E}_2^{(s)} = u_n^s G_{eve}^{(m)}.$$

Thus, the total side information will represent everything that will be achieved during the construction of $\tilde{E}_1^{(s)}$ and everything in the construction of $\tilde{E}_2^{(s)}$,

$$\tilde{E}^{(s)} = \tilde{E}_1^{(s)} \cup \tilde{E}_2^{(s)}.$$

So, the part of the $R$ random symbols from $\tilde{E}_1^{(s)}$ can be reconstructed in an identical way as the data reconstruction in the original MSR interference alignment code (Shah, 2012), and the part of the $R$ symbols from the failed node $\tilde{E}_2^{(s)}$, can be repaired in the same manner used for the repair process in the original MSR interference alignment code given in (Shah, 2012). Thus, this represents the first step of the information-theoretic secrecy proof for decoding all random symbols.

The number of chosen random symbols is $R = B - B^{(s)} = (l_1 + l_2)\alpha + (k - l_1 - l_2)l_2$. The number of observed symbols $\varepsilon$ can be calculated as the $l_1\alpha$ symbols obtained from the first observation plus the $(n - l_2)\beta$ symbols passed by each of the remaining $l_2$ nodes during the repair process in the second observation. In the second observation we need to subtract the already known symbols from the first observation, which are $\beta l_1$ symbols for each $l_2$ observation. Thus, the total number of observed symbols $\varepsilon$ is $l_1\alpha + (n - l_2)\beta l_2 - \beta l_1 l_2$. In our construction, for $n = 2k$ and $d = n - 1$, $\alpha = d - k + 1 = k$, $l_1 + l_2 = l$ and $\beta = 1$, so the expression becomes $(l_1 + l_2)\alpha + (k - l_1 - l_2)l_2$. We can say that the entropy of the random symbols and the entropy of the eavesdropped symbols are equal $H(\varepsilon) = H(R)$. This claim proves the second condition for achieving security in the storage system and that $H(\varepsilon) \leq H(R)$.

Last part of the proof establishes that the eavesdropper obtains no information about the message. In other words, the mutual information between all message symbols $U$ and total observed symbols $\varepsilon$ is zero,

$$\begin{aligned}
I(U;\varepsilon) &= H(\varepsilon) - H(\varepsilon \mid U) \\
&\overset{(a)}{\leq} H(R) - H(\varepsilon \mid U) \\
&\overset{(b)}{=} H(R) - H(\varepsilon \mid U) + H(\varepsilon \mid U, R) \\
&= H(R) - I(\varepsilon; R \mid U) \\
&= H(R) - (H(R \mid U) - H(R \mid \varepsilon, U)) \\
&\overset{(c)}{=} H(R) - H(R \mid U) \\
&\overset{(d)}{=} H(R) - H(R) \\
&= 0,
\end{aligned}$$

where $(a)$ follows from the result of Step 2, $H(\varepsilon) \leq H(R)$; $(b)$ follows since every symbol in the system is a function of $U$ and $R$ giving $H(\varepsilon \mid U, R) = 0$; $(c)$ follows from the result in Step 1; and $(d)$ follows since the random symbols are independent of the message symbols.

### 4.2 EXAMPLE OF SECURE CODE CONSTRUCTION

In this subsection we give a concrete example of secure Interference Alignment code in a presence of $l = l_1 + l_2$ eavesdroppers. With this example we prove the achievability of the general code construction and the level of security that is guaranteed with this scheme concept. For the example $[n=6, k=3, d=5]$ we have $\alpha = d-k+1 = 3$, and $B = k\alpha = 9$ ( $U = \{a_1, a_2, a_3, a_4, a_5, a_6, a_7, a_8, a_9\}$ ). The eavesdropper model is $(l_1, l_2) = (1,1)$. Based on that, using (6), the number of secure message symbols is $B^{(s)} = 2$. The number of random message symbols is $R = B - B^{(s)} = 9 - 2 = 7$. Regarding this, we must replace seven message symbols $a_1, a_2, a_3, a_4, a_5, a_6, a_7$ with random symbols $r_1, r_2, r_3, r_4, r_5, r_6, r_7$, drawn uniformly and independently from $\mathbb{F}_7$. After performing that, we are getting the new message with symbols $r_1, r_2, r_3, r_4, r_5, r_6, r_7, a_8, a_9$. For completing the process, we also need to define generator matrices. Because in our paper we are using the Interference Alignment concept, in this case we are distinguishing two types of generator matrices, which are explained in following.

1. Design of Systematic Generator Matrices:
   Knowing that $k=3$, the first $k$ nodes are systematic and the data stored on these nodes are uncoded. The generator matrices are given by,

$$G^{(1)} = \begin{bmatrix} I_3 \\ 0_3 \\ 0_3 \end{bmatrix}, \quad G^{(2)} = \begin{bmatrix} 0_3 \\ I_3 \\ 0_3 \end{bmatrix}, \quad G^{(3)} = \begin{bmatrix} 0_3 \\ 0_3 \\ I_3 \end{bmatrix}.$$

   where $0_3$ and $I_3$ are $(3 \times 3)$ zero matrix and identity matrix respectively.

2. Design of Parity Generator Matrices:
   The Cauchy matrix (Bernstein, 2005) is used for construction of the data that will be stored in the parity nodes, when interference alignment concept is employed. For this example, the matrix is a $(3 \times 3)$ matrix, such that each of the submatrices is a full rank matrix,

$$\Psi_3 = \begin{bmatrix} \psi_1^{(4)} & \psi_1^{(5)} & \psi_1^{(6)} \\ \psi_2^{(4)} & \psi_2^{(5)} & \psi_2^{(6)} \\ \psi_3^{(4)} & \psi_3^{(5)} & \psi_3^{(6)} \end{bmatrix}.$$

   The property of full rank is satisfied, because of the use of Cauchy matrix, that by its construction meets this property. The dimension of the matrix is 3 due to the number of parity nodes equal to $(n-k) = 6-3 = 3$.

   Based on that, the generator matrix of the parity nodes $m = 4,5,6$ is given by,

$$G^{(m)} = \begin{bmatrix} 2\psi_1^{(m)} & 0 & 0 \\ 2\psi_2^{(m)} & \psi_1^{(m)} & 0 \\ 2\psi_3^{(m)} & 0 & \psi_1^{(m)} \\ \psi_2^{(m)} & 2\psi_1^{(m)} & 0 \\ 0 & 2\psi_2^{(m)} & 0 \\ 0 & 2\psi_3^{(m)} & \psi_2^{(m)} \\ \psi_3^{(m)} & 0 & 2\psi_1^{(m)} \\ 0 & \psi_3^{(m)} & 2\psi_2^{(m)} \\ 0 & 0 & 2\psi_3^{(m)} \end{bmatrix}.$$

The generator matrix is designed regarding the functioning of the interference alignment method, with the necessary concern of performing the repair and the reconstruction process. An example of $[6,3,5]$ MSR code using the interference alignment method is given in Fig. 1, where the Cauchy matrix $\Psi$ is

$$\Psi = \begin{bmatrix} 5 & 4 & 1 \\ 2 & 5 & 4 \\ 3 & 2 & 5 \end{bmatrix}.$$

| Node 1 | $r_1$ | $r_2$ | $r_3$ |
|---|---|---|---|
| Node 2 | $r_4$ | $r_5$ | $r_6$ |
| Node 3 | $r_7$ | $a_8$ | $a_9$ |
| Node 4 | $3r_1+4r_2+6r_3$<br>$0+5r_2+0$<br>$0+0+5r_3$ | $2r_4+0+0$<br>$3r_4+4r_5+5r_6$<br>$0+0+2r_6$ | $3r_7+0+0$<br>$0+3a_8+0$<br>$3r_7+4a_8+6a_9$ |
| Node 5 | $r_1+3r_2+4r_3$<br>$0+4r_2+0$<br>$0+0+4r_3$ | $5r_4+0+0$<br>$r_4+3r_5+4r_6$<br>$0+0+5r_6$ | $2r_7+0+0$<br>$0+2a_8+0$<br>$r_7+3a_8+4a_9$ |
| Node 6 | $2r_1+r_2+3r_3$<br>$0+r_2+0$<br>$0+0+r_3$ | $4r_4+0+0$<br>$2r_4+r_5+3r_6$<br>$0+0+4r_6$ | $5r_7+0+0$<br>$0+5a_8+0$<br>$2r_7+a_8+3a_9$ |

*Figure 1: Example of $[6,3,5]$ MSR code using interference alignment method, with seven message symbols replaced with random variables.*

The multiplication between the modified original message, with seven replaced symbols, $u_n^s = [r_1\ r_2\ r_3\ r_4\ r_5\ r_6\ r_7\ a_8\ a_9]$ and the generator matrix $G^{(m)}$ defines the secure interference alignment MSR code depicted in Fig. 1,

$$C^{(s)} = u_n^s G^{(m)}.$$

Using the newly constructed secure MSR interference alignment code, we are able to prove the properties for satisfying the reconstruction, repair and security processes. For the first two processes the proof is given in (Shah, 2012) as is explaned in Theorem 1 and 2. And for the security process the proof is elaborated in the following. To be more clearer, the threat model in our example can observe the symbols stored on Node 1 shown in Fig. 1 and symbols downloaded during the repair process of Node 3. Meaning we are stating that even the eavesdropper have access to those symbols, she can not reveal the entire message.

In following the proof is given through an example with concrete numbers. Let the eavesdropper in the first observation have access to $l_1 k$ stored data in the system, or in our case access to Node 1, where $k = \alpha = 3$. This means she has access to $B_{s(1)}^{eve} = l_1 k = 1 \times 3 = 3$ symbols $\{r_1, r_2, r_3\}$ from total $B = k\alpha = 3 \times 3 = 9$, forming the message $u_s^{eve} = [r_1\ r_2\ r_3\ 0\ 0\ 0\ 0\ 0\ 0]$, in the $C^{(s)}$ code. The eavesdropper, knowing the coding scheme, from the first observation will get the following construction,

$$E_1^{(s)} = u_s^{eve} G^{(m)}.$$

Additionally, from the other type of observation is accessing the downloaded data equal to $B_{s(2)}^{eve} = d\beta l_2 = dl_2 = 5\times 1 = 5$ symbols during the repair process of Node 3, where $\beta = 1$ in our construction. By assuming that Node 3 fails in our example, each of the remaining $d = n-1 = 5$ nodes will pass its third column, so that the eavesdropper in the second observation gains the $\{r_3, r_6, r_7, a_8, a_9\}$ symbols and obtains the following columns during the repair process,

$$E_2^{(s)} = \begin{bmatrix} \underline{e}_3 & \underline{0} & \underline{0} & \psi_1^{(4)}\underline{e}_3 & \psi_1^{(5)}\underline{e}_3 & \psi_1^{(6)}\underline{e}_3 \\ \underline{0} & \underline{e}_3 & \underline{0} & \psi_2^{(4)}\underline{e}_3 & \psi_2^{(5)}\underline{e}_3 & \psi_2^{(6)}\underline{e}_3 \\ \underline{0} & \underline{0} & \underline{0} & \varepsilon\underline{\psi}^{(4)} & \varepsilon\underline{\psi}^{(5)} & \varepsilon\underline{\psi}^{(6)} \end{bmatrix},$$

where $\underline{e}_3$ denotes the third unit vector of length $\alpha$ and $\underline{0}$ denotes a zero vector of length $\alpha$. The element $\varepsilon$ from $F_q$ is $\varepsilon \neq 0$ and $\varepsilon^2 \neq 0$, by Definition 1 in (Shah, 2012) about Cauchy matrix, which is necessary condition for performing the reconstruction process.

Except for the third component in the example, every other component is aligned with the vector $\underline{e}_3$, which will show that some $\alpha = 3$ linear combinations of the columns above will give us a matrix whose third component equals the $(\alpha \times \alpha) = (3\times 3)$ identity matrix, and has zeros everywhere else. This is a consequence of the interference alignment structure in combination with the linear independence of the $\alpha = 3$ vectors in the desired component,

$$\{\underline{\psi}^{(4)}, \underline{\psi}^{(5)}, \underline{\psi}^{(6)}\},$$

where the $3$-th column is given by

$$\underline{\psi}^{(3)} = \begin{bmatrix} \psi_1^{(3)} \\ \vdots \\ \psi_3^{(3)} \end{bmatrix}.$$

The eavesdropper after both constructions can examine which symbols have revealed, they in this example are $\{r_1, r_2, r_3, r_3, r_6, r_7, a_8, a_9\}$. Based on the above given example we can conclude that some of the symbols are repeated, more specifically $r_3$ can be obtained either during the first or the second observation. Thus, the number of repeated symbols $h_r = 1$ should be subtracted from the total observed symbols, i.e., these symbols should not be taken into account and become $\{r_1, r_2, r_3, r_6, r_7, a_8, a_9\}$. Regarding this, the total number of message symbols observed by the eavesdropper is obtained as the sum of the two types of observations explained above,

$$B_s^{eve} = B_{s(1)}^{eve} + B_{s(2)}^{eve} - h_r = 3+5-1 = 7.$$

According to the Step 1 of proving the information-theoretic secrecy, we first need to show that for given message symbols as a side information, the eavesdropper can decode all the random symbols. Therefore, for the first type of observation we generate a new message vector $u_s^{eve}$ of length $k\alpha = 9$ consisting of only the symbols seen by the eavesdropper, and all other symbols zeros. This part of the side information is defined as,

$$\widetilde{E}_1^{(s)} = u_s^{eve} G^{(m)},$$

where $m = 1,2,3$ are the systematic nodes and $m = 4,5,6$ are the parity nodes.

In our example $l_1 = 1$, or more precise Node 1 is observed, meaning $u_s^{eve} = [r_1 \ r_2 \ r_3 \ 0\ 0\ 0\ 0\ 0\ 0]$ and $\widetilde{E}_1^{(s)}$ in the construction will have only information for these symbols. We are assuming that the eavesdropper knows the encoding scheme.

For the second type of observation, we take the message $u_n^s = [r_1 \ r_2 \ r_3 \ r_4 \ r_5 \ r_6 \ r_7 \ a_8 \ a_9]$ and define a new generator matrix $G_{eve}^{(m)}$. The matrices $G_{eve}^{(m)}$ for $m = 1,2,3$, systematic nodes, accordingly will have non zero elements on the places where the eavesdropper has access during the repair. In our example, because Node 3 fails the third column in every matrix except the failed one will have non zero element, and in the other zeros. Similarly will be performed for $G_{eve}^{(m)}$, $m = 4,5,6$ (these are the parity nodes, since we assume that the failed node is a systematic node) that corresponds to the $l_2$ nodes in which the eavesdropper can access the repair downloads. The Cauchy matrices in the generator matrix have nonzero columns only at the places where the eavesdropper has access (parts that will be downloaded during the repair according to which nodes have failed), or in our example all stored symbols consist in the third column so that the second part of the side information in the construction will be

$$\widetilde{E}_2^{(s)} = u_n^s G_{eve}^{(m)}.$$

From the above stated we can conclude that the total side information will represent everything that will be achieved during the construction $\widetilde{E}_1^{(s)}$ and everything in the construction $\widetilde{E}_2^{(s)}$, $\widetilde{E}^{(s)} = \widetilde{E}_1^{(s)} \cup \widetilde{E}_2^{(s)}$.

Therefore, the part of the $R$ random symbols from $\widetilde{E}_1^{(s)}$ can be reconstructed in an identical way as the data reconstruction in the original MSR interference alignment code (Shah, 2012). In this example, because Node 1 is affected, we will assume that the data collector will contact $k = 3$ nodes, from which two systematic (Node 2 and Node 3) nodes and one parity (Node 4) node. In this case he can obtained all symbols stored in Node 2 and Node 3 in uncoded form and proceeds to subtract their effect from the symbols in Node 4. To reconstruct all message symbols he must decode the message symbols $\{r_1, r_2, r_3\}$ that are encoded using the matrix $G_1^{(4)}$ given by

$$G_1^{(4)} = \begin{bmatrix} 2\psi_1^{(4)} & 0 & 0 \\ 2\psi_2^{(4)} & \psi_2^{(4)} & 0 \\ 2\psi_3^{(4)} & 0 & \psi_3^{(4)} \end{bmatrix}.$$

Since the lower-triangular matrix is nonsingular all entries in a Cauchy matrix are nonzero and the message symbols $\{r_1, r_2, r_3\}$ can be recovered by inverting $G_1^{(4)}$.

Moreover, the part of the $R$ symbols from the failed node $\widetilde{E}_2^{(s)}$, can be repaired in the same manner used for the repair process in the original MSR interference alignment code given in (Shah, 2012). Because in our example we repair systematic node (Node 3), the process of recovering the lost data is explained. We know that each node stores $\alpha = k$ symbols. To repair the systematic node $\ell = 3$, $1 \leq \ell = 3 \leq k = 3$, each of remaining nodes must pass their respective $\ell = 3$-th symbol (column). Thus, from nodes 4, 5, and 6, the replacement node will obtain

$$\begin{bmatrix} 0 \\ 0 \\ \psi_1^{(4)} \\ 0 \\ 0 \\ \psi_2^{(4)} \\ 2\psi_1^{(4)} \\ 2\psi_2^{(4)} \\ 2\psi_3^{(4)} \end{bmatrix}, \begin{bmatrix} 0 \\ 0 \\ \psi_1^{(5)} \\ 0 \\ 0 \\ \psi_2^{(5)} \\ 2\psi_1^{(5)} \\ 2\psi_2^{(5)} \\ 2\psi_3^{(5)} \end{bmatrix}, \begin{bmatrix} 0 \\ 0 \\ \psi_1^{(6)} \\ 0 \\ 0 \\ \psi(6)_2 \\ 2\psi_1^{(6)} \\ 2\psi_2^{(6)} \\ 2\psi_3^{(6)} \end{bmatrix}.$$

It should be noted that the desired third components are scaled version of the respective columns of the Cauchy matrix $\Psi_3$. The interference among the first and second components are aligned the vector $[1\,0\,0]^t$. Meaning that Node 1 and 2, which are systematic passes a single vector designed to cancel out the interference, or specifically pass the vectors

$$\begin{bmatrix} 1 \\ 0 \\ 0 \\ 0 \\ 0 \\ 0 \\ 0 \\ 0 \\ 0 \end{bmatrix}, \begin{bmatrix} 0 \\ 0 \\ 0 \\ 1 \\ 0 \\ 0 \\ 0 \\ 0 \\ 0 \end{bmatrix}.$$

So, after interference cancellation, the replacement Node 3, has access to the columns

$$\begin{bmatrix} 0_3 \\ 0_3 \\ 2\Psi_3 \end{bmatrix},$$

The desired component is the scaled Cauchy matrix $\Psi_3$. The recovery of the lost symbols now can be made by multiplying this matrix on the right by $\frac{1}{2}\Psi_3^{-1}$ to obtain

$$\begin{bmatrix} 0_3 \\ 0_3 \\ I_3 \end{bmatrix}.$$

This represents the first step of the information-theoretic secrecy proof for decoding all random symbols.

The number of chosen random symbols is $R = B - B^{(s)} = (l_1 + l_2)\alpha + (k - l_1 - l_2)l_2 = (1+1)3 + (3-1-1)1 = 7$. The number of observed symbols $\varepsilon$ can be calculated as the $l_1\alpha = 3$ symbols obtained from the first observation plus the $(n-l_2)\beta = 5$ symbols passed by each of the remaining $l_2 = 1$ nodes during the repair process in the second observation. In the second observation we need to subtract the already known symbols from the first observation, which are $\beta l_1 = 1$ symbols for each $l_2 = 1$ observation. Thus, the total number of observed symbols $\varepsilon$ is $l_1\alpha + (n-l_2)\beta l_2 - \beta l_1 l_2 = 3 + (6-1)1 - 1 = 7$. In our construction, for $n = 2k = 2\times 3 = 6$ and $d = n-1 = 6-1 = 5$, $\alpha = d-k+1 = k = 5-3+1 = 3$, $l_1 + l_2 = 1+1 = 2 = l$ and $\beta = 1$, so the expression becomes $(l_1 + l_2)\alpha + (k - l_1 - l_2)l_2 = (1+1)3 + (3-1-1)1 = 7$. We can say that the entropy of the random symbols and the entropy of the eavesdropped symbols are equal $H(\varepsilon) = H(R)$. This claim proves the second condition for achieving security in the storage system and that $H(\varepsilon) \leq H(R)$.

Last part of the proof establishes that the eavesdropper obtains no information about the message. In other words, the mutual information between all message symbols $U$ and total observed symbols $\varepsilon$ is zero, as is proved in subsection A.

### 4.3 ANALYSIS OF THE REPAIR BANDWIDTH AND SECRECY CAPACITY IN THE MSR CODES AND MSR INTERFERENCE ALIGNMENT CODES

In this subsection we are making analysis regarding the repair bandwidth and the secure message size that can be achieved when is used our code, MSR interference alignment, and the MSR regenerating code.

The repair bandwidth, when some node fail, is better to be smaller in order faster recovering of the lost data. The advantage of the MSR interference alignment code (Shah, 2012) is the efficient functioning of the repair process. If we compare it with the repair bandwidth of the MSR regenerating codes, shown in Fig 2., we can conclude that the MSR intereference alignment code gives smaller bandwidth compare to the MSR codes. The amount of the repair bandwidth in case of MSR regenerating codes is given by (2). The repair bandwidth, $\gamma = d\beta$, of the MSR interference alignment code depends of the code construction given by, $[n = 2k, k, d = n-1]$, $d = \alpha + k - 1$, $\beta = 1$ and $k = \alpha$.

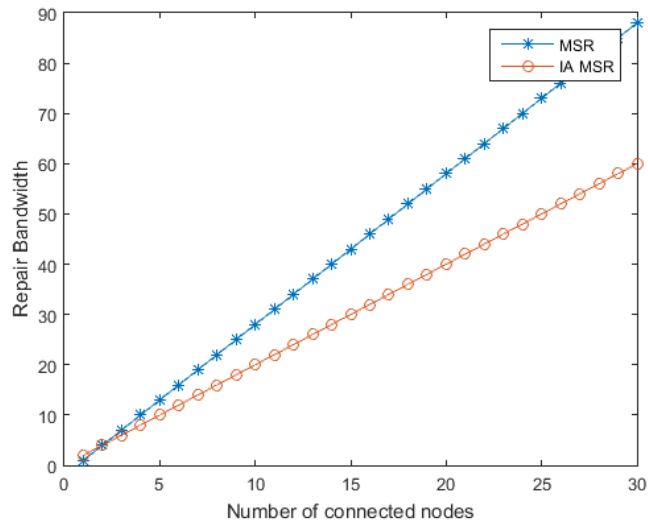

*Figure 2: Comparison of the repair bandwidth in the MSR regenerating codes and MSR interference alignment code for different number of contacted nodes up to $k=30$.*

The security capacity for our secure MSR interference alignment code in presence of passive eavesdroppers is given by (6). Additionally, we prove that our code is perfectly secure. The comparison between the secure MSR regenerating codes and the secure MSR interference alignment codes is shown in Fig 3. Here, we can notice that both lines are very close, but our code is slightly weaker. However, this small weakness can be compensated by the fact that our code gives better performance in the repair process and still is perfectly secure against the intruders.

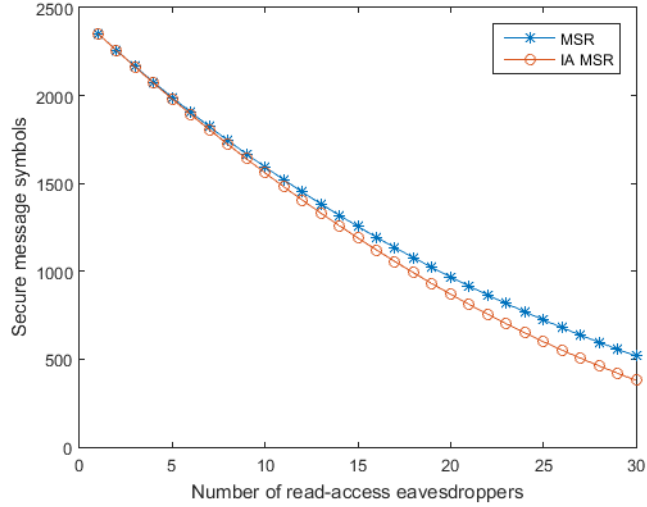

*Figure 3: Comparison of the size of the message that can be securely stored in the MSR regenerating codes and MSR interference alignment code for $k=30$ in presence of $l_1=1$ and $l_2=1,...,30$ compromised nodes.*

For the MSR codes, where the eavesdropper has an access to $l_1$ nodes, and listens $l_2$ nodes that are in the reparation process Goparaju et al. in (Goparaju, 2013) have established an upper bound of the achievable secure file size. So the secrecy capacity in this case is given by,

$$S^{(s)} = (k - l_1 - l_2)\left(1 - \frac{1}{d-k+1}\right)^{l_2} \alpha. \tag{14}$$

Therefore, the comparison shown in the figure for the secrecy capacities is done by making calculation among the equations (6) and (14).

## 5   CONCLUSION

This paper considers the security problem of constructing MSR interference alignment code. The MSR interference alignment code in (Shah, 2012) achieves the cut-set bound of repair bandwidth. Besides the optimal exact repair of systematic nodes, this explicit code is capable of performing data reconstruction. The main idea is that the construction is based on the interference alignment concept. This approach is used in the interference channels in wireless communications, where the method of canceling the interference of the other users resembles the repair process in DSS. Besides the reconstruction and repair capabilities of this code, here we show how to construct a new secure code based on the same principle which is resistant to attacks. The threat model is such that the eavesdropper can observe the data stored in any subset of the nodes, and the downloaded data during repair of another subset of failed nodes. The secure code design ensures that it restricts the information available to the eavesdropper and that the eavesdropper is unable to reveal the entire original message.